\newcommand{\vlowk}{V_{{\rm low}\,k}}
\documentclass[cits]{PoS}
\usepackage{amsmath}
\usepackage{amssymb}

\title{Neutron-rich Helium isotopes based on hyperspherical harmonics}

\ShortTitle{Helium isotopes from hyperspherical harmonics }

\author{\speaker{Sonia Bacca}\thanks{Thanks to my collaborators N. Barnea, R. Goerke and A. Schwenk for their help in obtaining these results.}\\
        TRIUMF, 4004 Wesbrook Mall,  Vancouver, BC V6T 2A3, Canada \\
        E-mail: \email{bacca@triumf.ca}}

\abstract{We present recent results for neutron-rich Helium isotopes obtained from the hyperspherical harmonics method. Ground-state properties, like the binding energy and the point-proton radius are shown for the two-neutron halo nucleus $^6$He using two-body low-momentum interactions derived from chiral forces. The applicability of the method to the four-neutron halo nucleus $^8$He is discussed. As an excited-state observable we present a recent calculation of the nuclear electric polarizability of $^6$He from a semi-realistic potential. A comparison of  the calculated quantities to experimental data is performed. }

\FullConference{The 7th International Workshop on Chiral Dynamics,\\
		August 6 -10, 2012\\
		Jefferson Lab, Newport News, Virginia, USA}

\begin{document}

\section{Introduction}

The physics of light neutron-rich nuclei is particularly interesting, because of the appearance of their exotic structures, like those of halo nuclei. The lightest nuclei of this kind are found in the Helium isotope chain: $^6$He as a two-neutron  and $^8$He as a four-neutron halo nucleus. They are both radioactive and undergo $\beta$-decay with a half life of $t_{1/2}=0.8$s and $0.1$s, respectively. Despite the short life time, a combination of atomic and nuclear physics techniques, has enabled precise measurements of ground-state observables like the energy and the charge radius  \cite{Maxime}. Excited-state properties, like electromagnetic transitions in the continuum have been investigated in the past with Coulomb dissociation experiments by Aumann {\it et al}.~\cite{Aumann}.
Tackling the theoretical study of these nuclei is very challenging, because one needs to simultaneously describe
the small separation energy of the halo neutrons and the large radius of the whole system.  Because they are light-mass nuclei, one can use  {\it ab-initio} techniques to study them.    Here, we will discuss the recent results obtained using the hyperspherical harmonics (HH) method.
  
 A major breakthrough in nuclear physics has been the development of chiral effective field theory, which is well routed to Quantum Chromo Dynamics. Even though several light nuclei have been investigated with chiral potentials, we are still missing a prediction of Helium halo nuclei from chiral Hamiltonians. Here, we show a first step taken in this direction by using low-momentum chiral two-nucleon forces.
  
The paper is organized as follows. In section 2 we will introduce the hyperspherical harmonics method. In section 3 and 4 we will present results for ground-state and excited-state properties, respectively.
Finally, in section 5 we will draw some conclusions.

\section{Hyperspherical Harmonics}

Given the Hamiltonian $H$ we use the HH expansion to solve the
Schr\"{o}dinger equation.  The HH method is typically a
few-body method used for 3 and 4-body systems.  
Using the powerful antisymmetrization
algorithm introduced in~\cite{Nir}, it is possible to extend the method to a larger mass number and tackle 
Helium halo nuclei.
The HH approach starts from the Jacobi coordinates
\begin{equation}
\label{jacobc}
\boldsymbol {\eta }_{0~~}=\frac{1}{\sqrt{A}}\sum _{i=1}^{A}\mathbf{r}_{i}\,,
\qquad \boldsymbol {\eta }_{k-1}=\sqrt{\frac{k-1}{k}}\left( \mathbf{r}_{k}-\frac{1}{k-1}\sum _{i=1}^{k-1}\mathbf{r}_{i}\right),\, k=2,...,A\,,
\end{equation}
where $\mathbf{r}_i$ are the particle coordinates.  Using the
$\boldsymbol {\eta }_{i}$ one can then transform to hyperspherical
coordinates composed of one hyperradial coordinate
$\rho=\sqrt{\sum_{i=1}^{A-1} \boldsymbol {\eta }_{i}^2}$ and a set of
$(3A-4)$ angles that we denote with $\Omega$ (for more details
see~\cite{Nir}). Using this coordinates one can recursively construct
the hyperspherical harmonics $\mathcal{Y}_{[K]}$ and use them as a complete basis to expand
the wave function. Such expansion reads
\begin{equation}
\label{expans}
\Psi( \boldsymbol {\eta }_{1}, ..., \boldsymbol {\eta }_{A-1},
  s_1,...,s_A, t_1,...,t_A)= \sum_{n}^{n_{\rm max}}  \sum_{[K]}^{K_{\rm max}}C_{[K] n} \, R_{n}(\rho) \,
{\cal Y}_{[K]}(\Omega,s_1,...,s_A, t_1, ...,t_A ),
\end{equation}
where $s_i$ and $t_i$ are the spin and isospin of the nucleon i,
respectively; $C_{[K] n}$ is the coefficient of the expansion, labeled
by $[K]$, which represents a cumulative quantum number that includes
the grandangular momentum $K$;  $n$ labels the hyperradial wave function
$R_{~n}(\rho)$.  The latter is  expanded in terms of 
the generalized Laguerre polynomials times an exponentially 
 falling off function, which is essential
to speed up the convergence of  halo nuclei, due to their extended 
tail. 
To further increase the convergence rate of
the calculations, we typically employ an effective interaction
in the 
hyperspherical harmonics (EIHH), as first introduced
in~\cite{EIHHlocal}.

\section{Results for ground state properties}

In the following, we present our results for
the ground state energy and the point-proton radius calculated using the hyperspherical 
harmonics method. As input Hamiltonian we employ a class of low-momentum potentials, which
are obtained applying a $\vlowk$ procedure \cite{Vlowk} on a starting chiral 
two-body potential \cite{EM_pot} at next-to-next-to-next-to leading order (N~$^3$LO).

\begin{figure}
\includegraphics[scale=0.37,clip=]{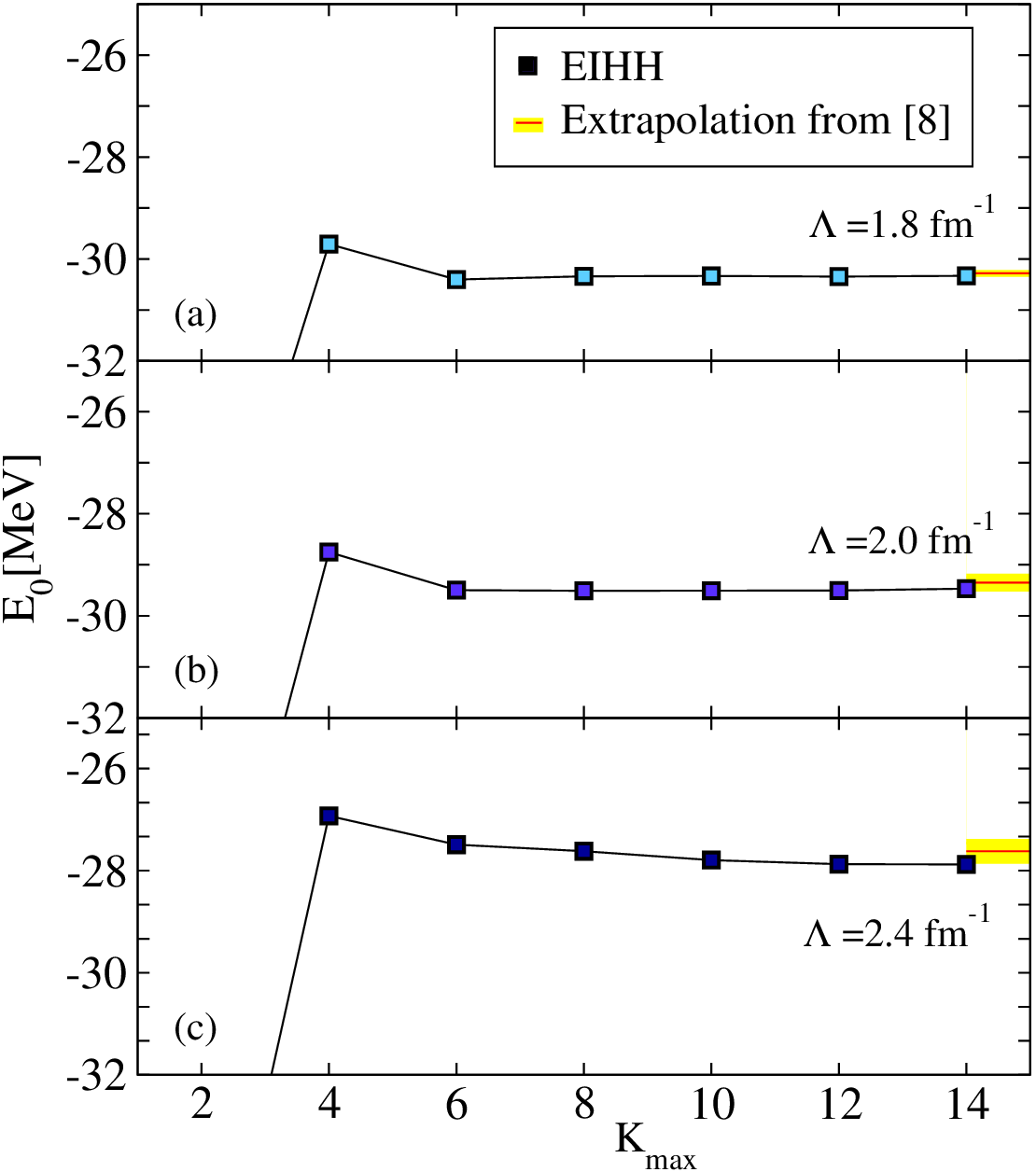}
\includegraphics[scale=0.37,clip=]{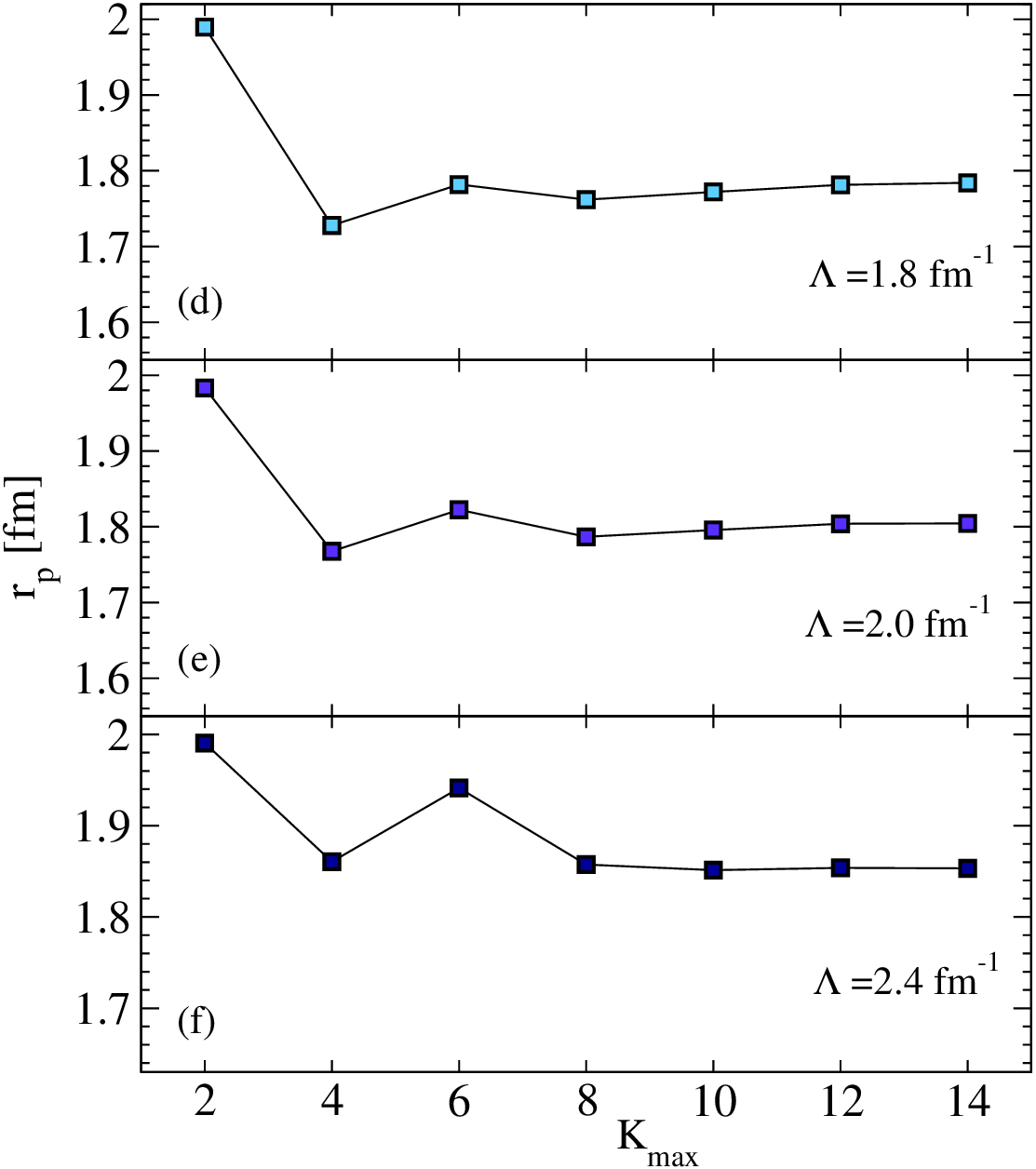}
\caption{(Color online) The $^6$He ground-state energy (left) and the point-proton radius (right) as  functions of $K_{\rm max}$ obtained for three different
  values of the cutoff $\Lambda=1.8,2.0,2.4$ fm$^{-1}$ of the $\vlowk$
  chiral potential.}
\label{fig_enrg_radii}
\end{figure}

In Fig.~\ref{fig_enrg_radii}, we
show the $^6$He convergence patterns of the EIHH method for the
ground-state energy and the point-proton radius as a function of the maximal grandangular momentum~\cite{Sonia_PRC}.  Three different cutoffs
$\Lambda=1.8,2.0,2.4$ fm$^{-1}$ of the $\vlowk$ chiral potential are shown. 
One observes that the convergence is very nice for all $\Lambda$'$s$, even for the largest cutoff.
Concerning the energy we also show that the extrapolated results from previous work on hyperspherical harmonics~\cite{Sonia_EPJ} agree nicely with EIHH.
The convergence rate is very good also for the point-proton radius, which allows us to provide solid results for ~$^6$He.

\begin{figure}
\begin{center}
\includegraphics[scale=0.31,clip=]{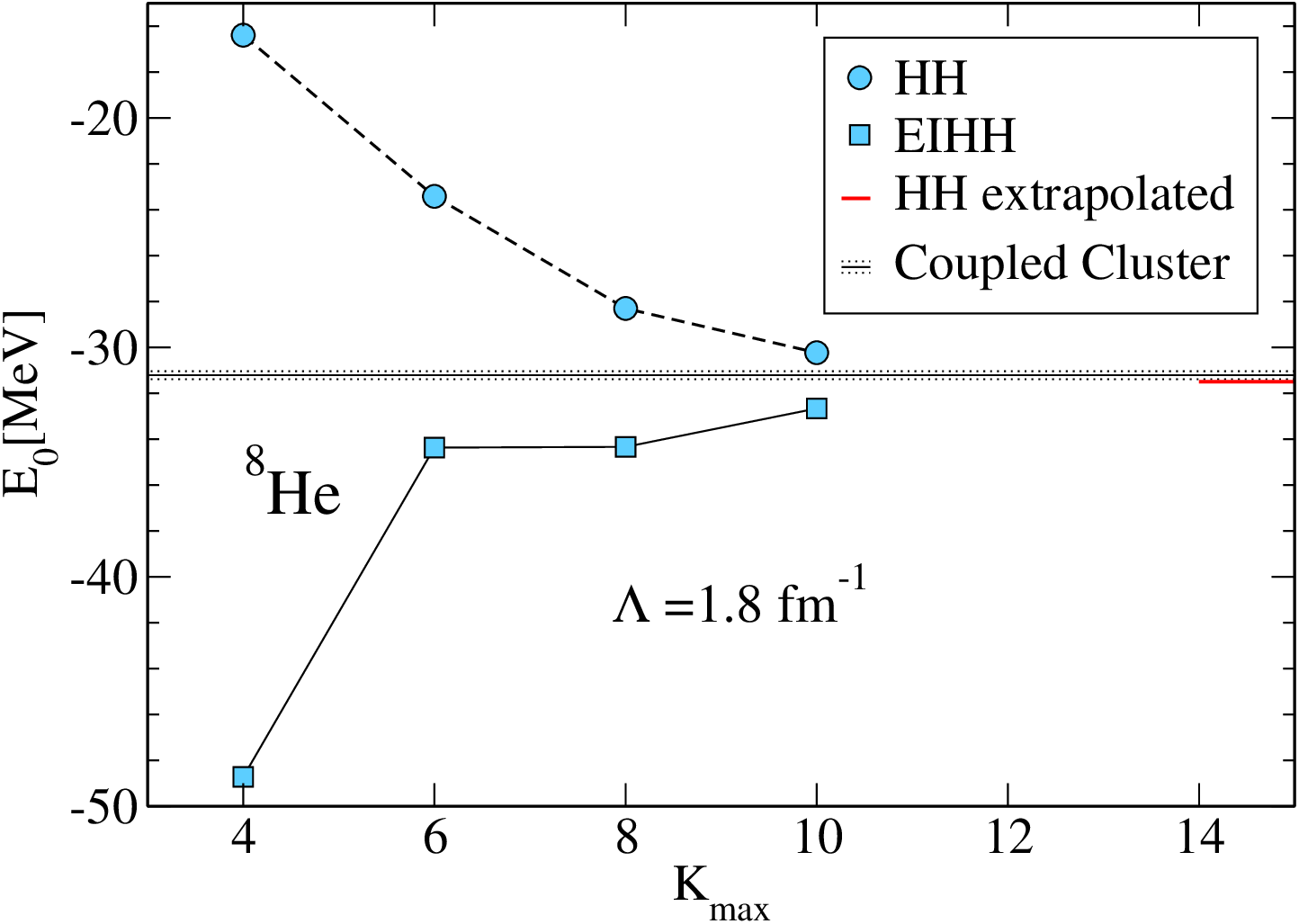}
\caption{(Color online)
The $^8$He ground-state energy calculated with the
  HH and EIHH methods as a function of 
  $K_{\rm max}$ obtained for a cutoff of $\Lambda=1.8$ fm$^{-1}$ of
  the $\vlowk$ chiral potential. The extrapolated HH results are shown
  as a reference and compared to the Coupled-Cluster results. }
\label{he8}
\end{center}
\end{figure}

We have also explored the four-neutron halo nucleus $^8$He with hyperspherical harmonics.
In Fig.~\ref{he8}, we show the $^8$He ground state energy
from a $\vlowk$ chiral potential with $\Lambda=1.8$
fm$^{-1}$ as a function of  $K_{\rm max }$. We present both the variational 
HH expansion (where we do not apply the effective interaction) and the EIHH results.
The convergence for ~$^8$He is quite slow, as indicated by the fact that  the HH and EIHH patterns do not merge yet at $K_{~\rm max}=10$. 
Interestingly, an extrapolation of the variational HH data 
lies close to the Coupled-Cluster result from~\cite{Sonia_EPJ} with the same interaction. 
Nevertheless, at the moment we are not able to provide precise results for ~$^8$He from hyperspherical
harmonics.
\begin{figure}
\begin{center}
\includegraphics[scale=0.31,clip=]{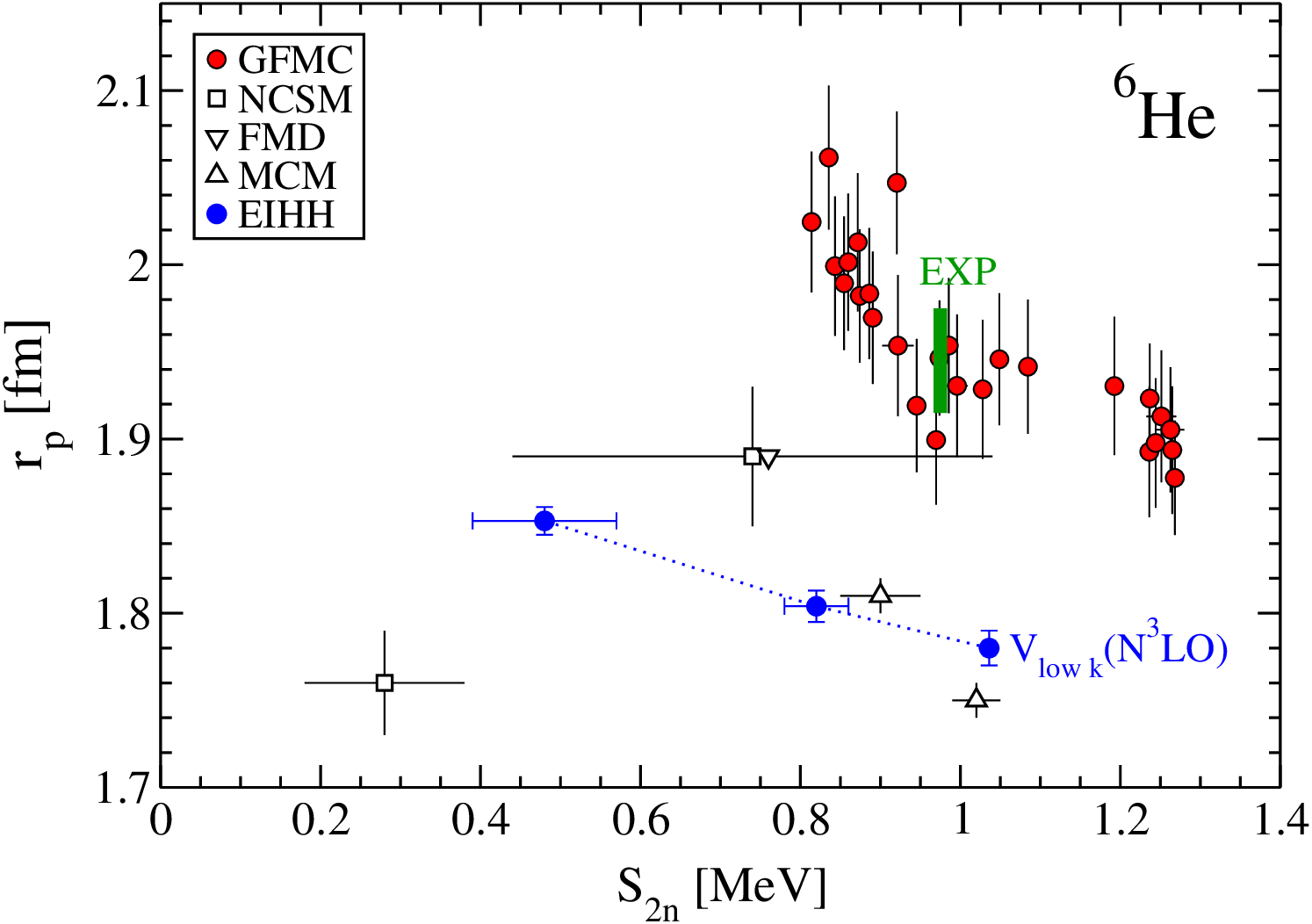}
\caption{(Color online) Correlation plot of the $^6$He  point-proton radius  versus the two-neutron separation
  energy $S_{2n}$. The experimental range is compared to theory based on different {\it ab-initio} methods (see text).} 
\label{He6_correlation}
\end{center}
\end{figure}
Thus, we concentrate on analyzing the correlation between energy and radius just for $^6$He.
In Fig.~\ref{He6_correlation},
we plot $r_{\rm p}$  versus the two-neutron separation energy $S_{~2n}$ and
  present a combined comparison of our results to experiment and other {\it ab-initio} calculations: Green's Function Monte Carlo (GFMC),
No Core Shell Model (NCSM),
Fermionic Molecular Dynamics (FMD) 
and Microscopic Cluster Model (MCM) (see also \cite{Sonia_PRC} and references therein).
The cutoff dependence of our  results with $\vlowk$ 
 allows us to study the correlation between these
observables: the radius increases as the separation energy decreases.
Our calculations do not reproduce simultaneously $r_{\rm ~p}$ and $S_{2n}$:
there exists an optimal value of $\Lambda$ where $S_{2n}$ is predicted in accordance with experiment, but $r_{\rm p}$ is not reproduced and vice-versa. 
Also, we would like to note that  other calculations which omit $3NF$, (all except from the GFMC)
do not go though the experimental band. This points  towards the importance of including three-nucleon forces
in the Hamiltonian.

\section{Results for excited state properties}

As an example of excited-state properties of halo nuclei we report about our recent calculation of the electric dipole polarizability $\alpha_E$ of $^6$He \cite{Goerke}. $\alpha_E$ is  related to the inelastic response of the nucleus to an externally applied electric field and is relevant
in the extraction of nuclear quantities from atomic spectroscopic
measurements.  The atomic energy levels, in fact, are affected by  polarization of the
nucleus due to the electric field of the surrounding electrons.
 The polarizability of ~$^6$He could be extracted from Coulomb dissociation measurement of the dipole transition by
Aumann {\it et al.}~\cite{Aumann} and was reported in Ref.~\cite{Moro} to be much bigger than
the polarizability of ~$^4$He.
\begin{figure}
\begin{center}
\includegraphics[scale=0.28,clip=]{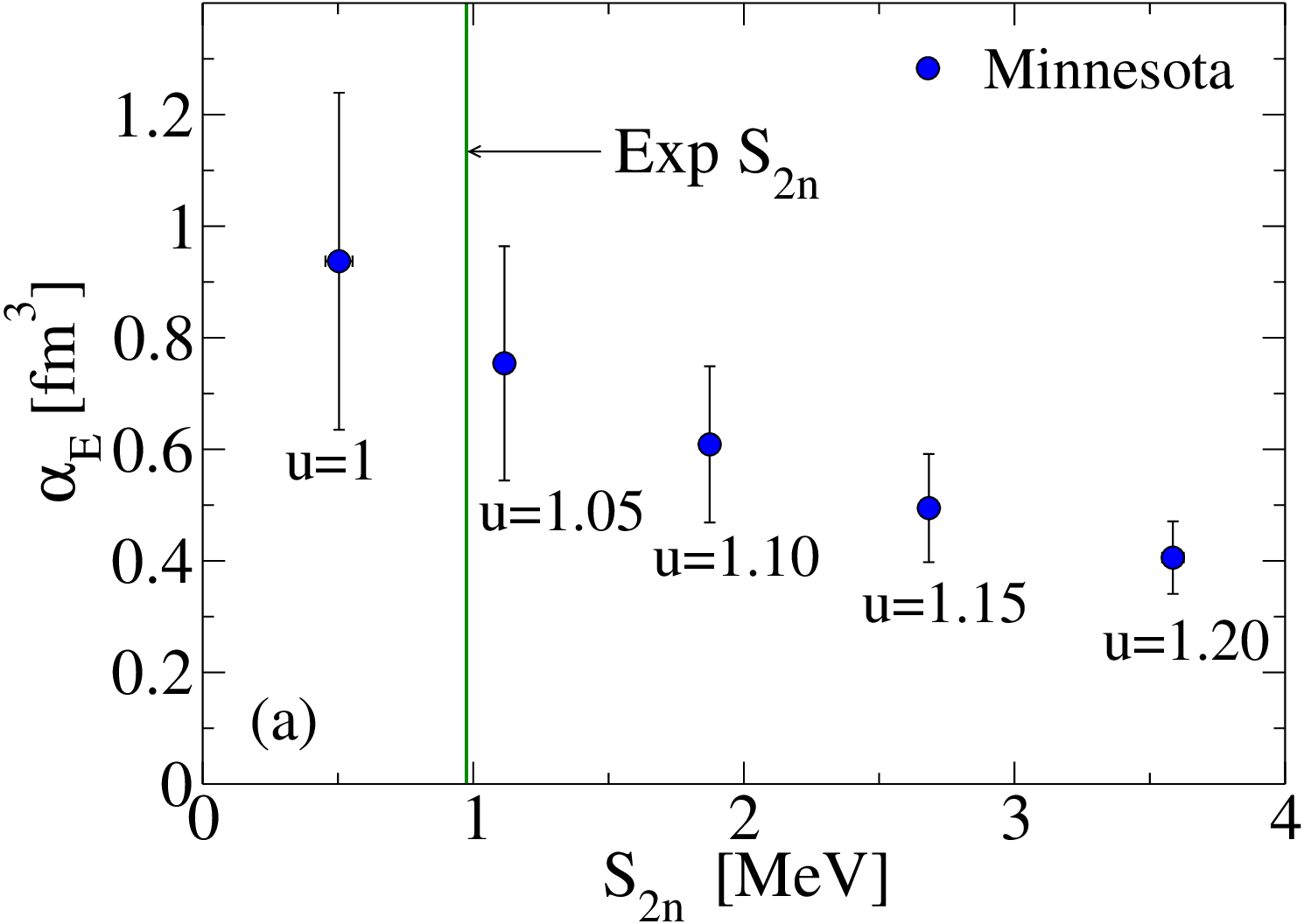}
\includegraphics[scale=0.28,clip=]{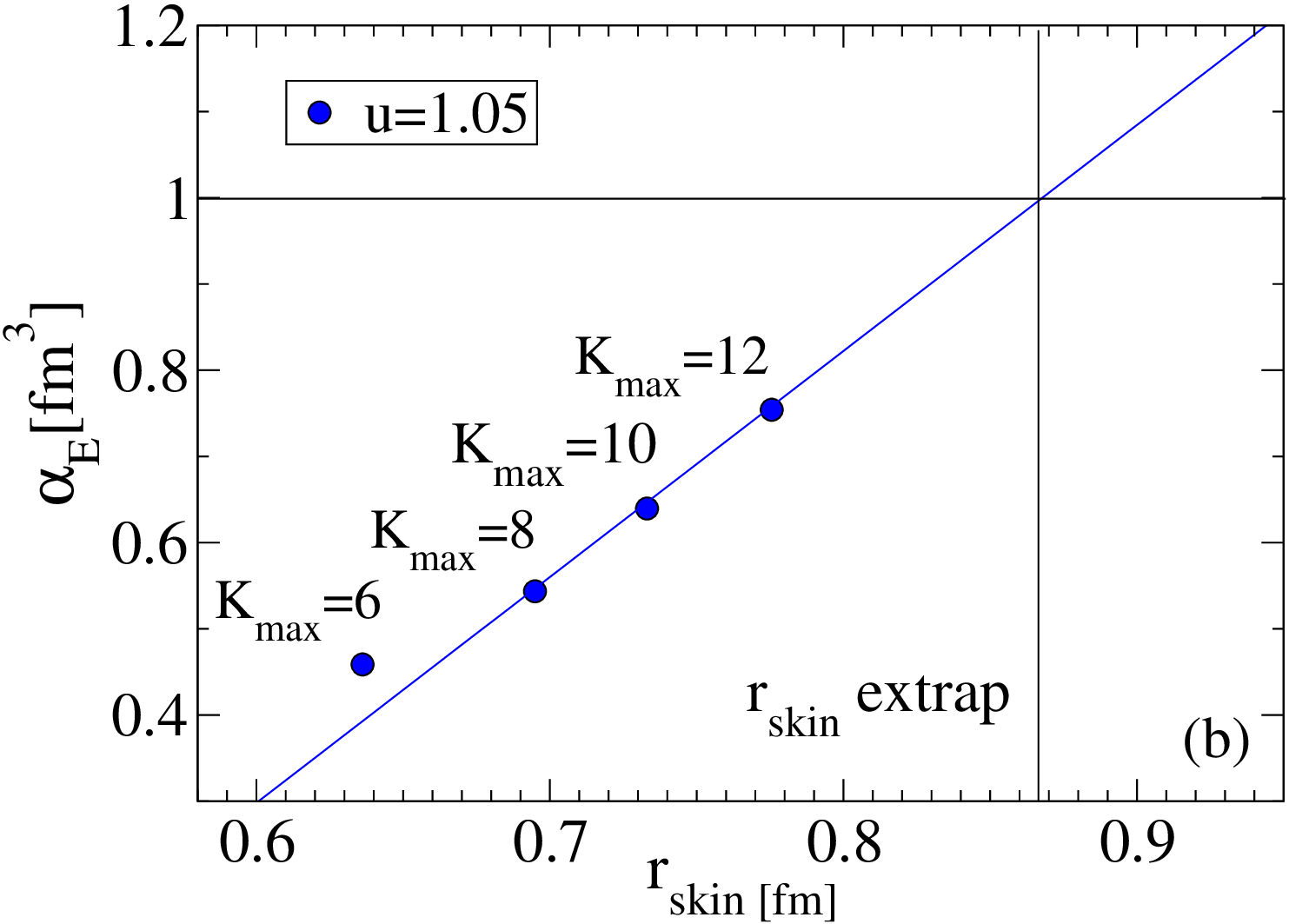}
\caption{(Color online) Panel (a): The correlation between $\alpha_E$ and
 $S_{2n}$ in $^6$He obtained with
the the Minnesota potential varying the parameter $u$. 
Panel (b): The correlation between  $r_{\rm skin}$  and $\alpha_E$ in different model spaces (different $K_{\rm max}$). Shown is also
the extrapolated value of $r_{\rm skin}$, which is used to estimate $\alpha_E$. 
} 
\label{fig_pol}
\end{center}
\end{figure}
The electric dipole polarizability
is defined by
\begin{equation}
\alpha_E=2 \alpha \sum_{f\ne0} \frac{|\langle \Psi_f| E1| \Psi_0\rangle|^2}{E_f -E_0}\,,
\label{pol}
\end{equation}
where $|\Psi_{0/f}\rangle$ is the ground state and final state of the nucleus and $E1$ is the dipole operator.
Because it requires the knowledge of the dipole spectrum of the nucleus, its theoretical evaluation is more involved
than a bound-state calculation. We perform our calculation with the EIHH method by using the Lanczos algorithm
with a starting dipole pivot as explained in \cite{Goerke}.
As nuclear potential we chose the  simple semi-realistic  Minnesota force, which reproduces the experimental value of the polarizability of ~$^4$He reasonably well. Within this force model
 we can add attractive  $P-$wave interactions  by changing the parameter $u$ (see \cite{Goerke} for details). 
This mostly affects ~$^6$He, without substantially changing $^4$He.
By varying $u$ we first observe a correlation of $\alpha_E$ vs $S_{2n}$, as shown in Fig.~\ref{fig_pol}(a).
We have chosen $u$ so that the halo feature, represented by $S_{2n}$, is reproduced.
We then study the correlations between $\alpha_E$ and the skin radius $r_{\rm skin}=r_n -r_p$, 
where $r_n$ is the mean point-neutron radius.
By varying the model space we observed that 
 $\alpha_E$ and $r_{\rm skin}$ are correlated linearly for  $K_{\rm max} \geq 6$  as
$\alpha_E=a+b~r_{\rm skin}$.
From our theoretical data we fit  the coefficients $a$ and $b$
and then we used them to estimate the polarizability
out of a bound-state calculation of the skin radius.
The calculation of $r_{\rm ~skin}$, in fact, does not require an expansion on the dipole excited states and as such is less computationally demanding and can be performed for larger model spaces ($K_{~\rm max}=16$) and then extrapolated exponentially, leading to
 $r_{\rm skin}=0.87(5)$ fm. 
Using our extrapolated skin radius and the linear dependence,  we estimate the theoretical nuclear electric polarizability of ~$^6$He to be  
$\alpha_E=1.00(14)$ fm$^3$. The error bar is obtained by propagating the errors on $a$, $b$ and $r_{\rm skin}$.
We observe that our theoretical estimate is about a factor of two smaller than the experimental value of $\alpha_E^{\rm exp}=1.99(40)$ fm$^3$ \cite{Moro}. This points toward a potential disagreement between theory and experiment. Investigations with chiral potentials can possibly help understanding this discrepancy.
\section{Conclusions}
In conclusion, we have presented our recent results on Helium halo nuclei from hyperspherical harmonics. 
The binding energy and the radius can be precisely calculated for ~$^6$He using chiral low-momentum  two-body forces. The obtained cutoff dependence together with a comparison to the experiment serves to highlight the importance of three-nucleon forces. We also discussed the nuclear dipole polarizability of ~$^6$He as an excited state observable, which
 has recently attracted  attention. Our estimate from simplified nuclear potentials leads to a disagreement with experimental data, which will be hopefully clarified in the future when realistic chiral potentials will be used.

\end{document}